\documentclass[aps,twocolumn,pre,showpacs,superscriptaddress,groupedaddress,amsmath,amssymb]{revtex4-1}
\usepackage{epsfig}
\usepackage{graphics}
\usepackage{lipsum}
\usepackage{sidecap}
\usepackage{graphicx} 
\usepackage{dcolumn} 
\usepackage{bm}
\usepackage{xcolor,colortbl}
\usepackage{color, colortbl}
\definecolor{darkgray}{rgb}{0.66, 0.66, 0.66}
\definecolor{yellow-green}{rgb}{0.6, 0.8, 0.2}
\definecolor{deeppink}{rgb}{1.0, 0.08, 0.58}
\definecolor{darkviolet}{rgb}{0.58, 0.0, 0.83}
\definecolor{darkcyan}{rgb}{0.0, 0.55, 0.55}

\usepackage{sidecap}  
\begin{document}
\preprint{APS/123-QED}

\title{Phase-flip chimera induced by environmental nonlocal coupling}

\author{V.~K.~Chandrasekar$^{1}$}
\author{R.~Gopal$^{2,3}$}
\author{D.~V.~Senthilkumar$^{4}$}
\author{M.~Lakshmanan$^2$}
 
\affiliation{
$^{1}$Centre for Nonlinear Science \& Engineering, School of Electrical \& Electronics Engineering, SASTRA University, Thanjavur- 613 401, India.\\
$^{2}$Centre for Nonlinear Dynamics, School of Physics, Bharathidasan University, Tiruchirapalli-620024, India.\\
$^{3}$Department of Physics, Nehru Memorial College, Puthanampatti,
Tiruchirapalli 621 007, India.\\
$^{4}$ School of Physics, Indian Institute of Science Education and Research, Thiruvananthapuram-695016, India.
}
\date{\today}
            
\begin{abstract}
We report the emergence of a  collective dynamical state, namely phase-flip chimera,
from an ensemble of identical nonlinear oscillators that are coupled indirectly
via the dynamical variables from a common environment, which in turn 
are nonlocally coupled.  The phase-flip chimera is characterized by the coexistence of two adjacent
out-of-phase synchronized coherent domains interspersed by an incoherent domain,
in which the nearby oscillators are in out-of-phase synchronized 
states. Attractors of the coherent domains are either from the same or different basins of attractions depending on whether they are periodic or chaotic. Conventional chimera precedes the phase-flip chimera in general.
Further, the phase-flip chimera emerges after the completely synchronized evolution of the ensemble 
in contrast to conventional chimeras which emerge as an intermediate
state between completely incoherent and coherent states.  We have also characterized the observed dynamical transitions
using the strength of incoherence, probability distribution of correlation coefficient and the framework of master stability function.
\end{abstract}


\maketitle

\section{Introduction}

Identification of an intriguing collective dynamical state,
namely the chimera state, in an ensemble of coupled identical nonlinear oscillators 
with nonlocal coupling~\cite{Kuramoto:84,Pikovsky:01,Winfree:01,Kuramoto:1}
has initiated intense research activities in the recent 
literature~\cite{Kuramoto:1,laing2000,ioym2011,sheeba09,uji2013,chan15}. A chimera state
represents a spatially inhomogeneous state characterized by coexisting coherent and incoherent domains 
in an ensemble of identical oscillators. 
Experimentally, chimera has also been revealed in populations of coupled  
chemical oscillators~\cite{tinsley2012}, in electro-optical systems~\cite{Hage2012} and in 
metronomes~\cite{martens2013}. Real world examples mimicking chimera states can be found in
power grids~\cite{Filatrella2008}, in unihemispheric sleep of animals~\cite{rottenberg2000},
in multiple time scales of sleep dynamics~\cite{Olb11}, etc. 
Different types of chimera states such as amplitude mediated chimera~\cite{Sethia:13}, 
intensity induced chimera~\cite{Gopal1:14},  and so on have also been identified~\cite{Sethia1:14}. 
Recently, it has been 
shown that a symmetry breaking coupling in the Stuart-Landau oscillators leads to 
the manifestation of chimera death~\cite{zak2014,prema:15}. 
 The collective state with inhomogeneous flipping
between the steady states (oscillation death) of an ensemble  was called  chimera death.  
Very recently, noise  induced coherence-resonance chimeras in  a network of excitable elements
was reported~ \cite{Seme2015}.  Coherence-resonance chimeras
are associated with alternating switching of the location of coherent and incoherent domains.

%
\begin{figure*}
\centering
\includegraphics[width=2.0\columnwidth]{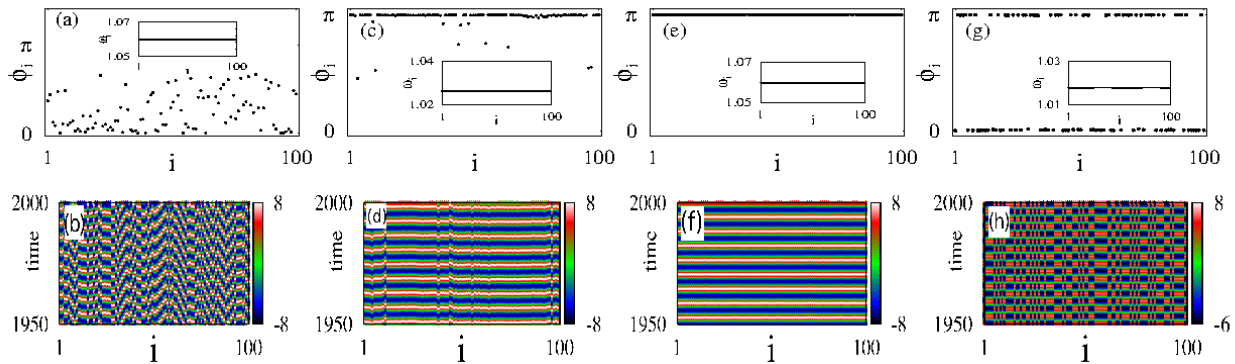}
\caption{(Color online) Snapshots of the instantaneous phases $\phi_{i}$ (top row) and the space-time evolution
 (bottom row) of the ensemble of R\"ossler oscillators in the periodic regime for the coupling radius $r=0.3$ of the 
nonlocal coupling and for different values of the strength of the nonlocal coupling exhibiting 
(a)-(b) desynchronized state for $\varepsilon=0.02$,
(c)-(d) chimera state for $\varepsilon=0.08$, 
(e)-(f) synchronized state for $\varepsilon=0.5$  and 
(g)-(h) phase-flip chimera for $\varepsilon=1.5$. 
Time averaged frequencies of all the oscillators are shown in the insets. f  Other parameter values are $a=0.165$, $b=0.4$, $c=8.5$ and $\alpha=1$.}
\label{pre_fig1}
\end{figure*}
 In this manuscript, we unravel a novel dynamical regime emerging from  the
inhomogeneous synchronized states. 
In particular, we consider an ensemble of identical  nonlinear oscillators coupled
via a common dynamic environment with nonlocal coupling.  We show that
the ensemble of oscillators splits into coexisting coherent and incoherent domains 
for appropriate strength of the nonlocal coupling. 
The nearby oscillators in the coherent domains exhibit in-phase synchronized oscillations
while the  adjacent oscillators in the incoherent domains exhibit out-of-phase
oscillations. 
It is to be noted that the term coherent/incoherent domains here represents the
homogenous/inhomogenous nature of the dynamics of the attractors in the corresponding domain and 
does not refer to the nature of the attractors, that is whether they are coherent attractors with a 
fixed center of rotation or incoherent
attractors with more than one center of rotation.

Further, we find that some of the nearby coherent domains 
exhibit in-phase oscillations in one of the domains and  anti-phase oscillations in the other resembling the
phase-flip bifurcation/transition~\cite{Prasad:05}. Specifically, the out-of-phase synchronized 
nearby coherent domains are interspersed by an incoherent domain (where the phases of the adjacent oscillators flip between 0 and $\pi$) at the phase-flip 
transition, which  we call a phase-flip chimera.  
It is a dynamically active emerging behavior
in contrast to the chimera death~\cite{zak2014,prema:15}, where nearby oscillators populate  the  same   
branch of the inhomogeneous steady state in the coherent domain while the nearby oscillators populate  different branches of the inhomogeneous steady state in the incoherent domain.
Further, we find that the conventional chimera is preceded by the phase-flip chimera
in addition to the other collective dynamical regimes such as coherent and synchronized states
in an ensemble of the paradigmatic R\"ossler oscillators both in the periodic and chaotic
regimes.  
We have used the measures, namely the strength of incoherence and the probability distribution of the
correlation coefficient to characterize the phase-flip chimeras and the observed dynamical transitions.
In addition, we have also employed the framework of the master-stability function (MSF) to demarcate the synchronized and desynchronized
parameter space which agrees very well with the simulation results. It is to be pointed out that the
synchronized regime is a multistability regime coexisting with conventional chimeras and phase-flip chimeras depending on the distribution of the initial conditions.

 The plan of the paper is as follows. We discuss the emergence of phase-flip chimera 
in an ensemble of R\"ossler oscillators in the periodic regime in Sec. II and in the chaotic regime in Sec. III.
We provide certain quantification measures such as the strength of incoherence, probability distribution
of the correlation coefficient and the framework of the master stability functon to characterize the 
different collective dynamical behavior in the ensemble of oscillators in Sec. IV.  Discussion on
the global dynamical behavior in terms  of two-parameter phase diagrams is provided in Sec. V. Emergence
of phase-flip chimera with global coupling among the agents in the common environment  will be
discussed in Sec. VI and conclusion will be provided in Sec. VII. In Appendix A, we consider the effect of relaxation time while the effect of coupling to other variables is discussed in Appendix B.

\section{Phase-flip chimera in an ensemble of R\"ossler oscillators in the periodic regime}
To elucidate the above results, we consider an ensemble of identical R\"ossler oscillators
with a common dynamic environmental coupling represented as
\begin{subequations}
\begin{align}
\dot{x}_{i}=&\,-y_{i}-z_{i},\\
\dot{y}_{i}=&\,x_{i}+ay_{i},\\
\dot{z}_{i}=&\,b+z_{i}(x_{i}-c)+kw_{i}, \\ 
\dot{w}_{i}=&\,-\alpha w_{i}+\frac{z_{i}}{2}+\frac{\varepsilon}{2P}\sum_{j=i-P}^{j=i+P}(w_{j}-w_{i}), \label{ross4}
\end{align}
\label{ross}
\end{subequations}
$i=1,\ldots,N$, where $a$, $b$, $c$ and $\alpha$ are the system parameters, 
$N$ is the number of oscillators in the ensemble. The oscillators in the 
ensemble are coupled indirectly via nonlocally coupled dynamic agents $w_i$
in the common environment. $k$ is the strength with which
the agent $w_{i}$ from the common environment interacts with the $i$th oscillator
in the ensemble. $\varepsilon$ is the strength of the nonlocal coupling.
$P\in[1,N/2]$ is the number of nearest neighbors on each side of any oscillator 
in the ring with a coupling radius $r=\frac{P}{N}$. The environment/medium
plays a crucial role in facilitating the complex collective  dynamics such as
decoherence, dissipation and relaxation in quantum systems~\cite{Ammann1998},
in coordinated rhythms in biological systems~\cite{kuz2004}, 
and in quorum sensing~\cite{li2012}.  The dynamics of the individual agent given by $\dot{w}_{i}$
in Eq.~(\ref{ross}d) is related to the interactions of molecules between the cells
and their environment~\cite{Ullner:07,Ambi:10,Amit:12}. 
In the following, we will demonstrate
the existence of phase-flip chimera in an ensemble of R\"ossler oscillators in
both periodic and chaotic regimes.

\begin{figure}
\centering
\includegraphics[width=1.0\columnwidth]{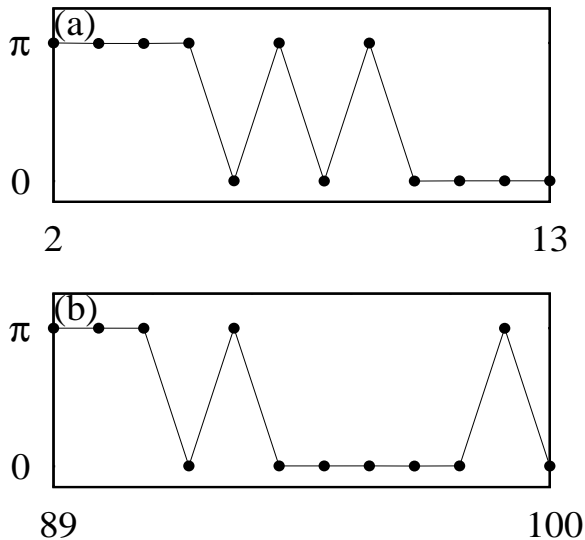}
\caption{(Color online) Enlarged regions of Fig.~\ref{pre_fig1}(g) to clearly show the
phase-flip chimera characterized by coexisting coherent and incoherent domains, where the
nearby oscillators exhibit in-phase oscillations in the coherent domain while the adjacent 
oscillators exhibit out-of-phase oscillations in the incoherent domain.}
\label{pre_fig2}
\end{figure}
Individual R\"ossler oscillators in the ensemble exhibit periodic oscillations
for the parameters $a=0.165$, $b=0.4$ and $c=8.5$.  We have used random initial conditions 
uniformly  distributed between $-1$ to $+1$. We have fixed 
$N=100$, $k=10$, and the coupling radius $r$ 
as $r=0.3$  ($N=100$ is chosen  only for clarity in the sense that the jumping phases 
can be distinctly seen. One can indeed choose any value of $N$ for our analysis). Snapshots of the instantaneous
phases $\displaystyle \phi_i=\arctan(y_i/x_i)$, $i=1,2...N$, and the spatiotemporal evolution of the oscillators are depicted in
Fig.~\ref{pre_fig1} for different values of the strength of the nonlocal coupling
$\varepsilon$.  The oscillators evolve in asynchrony  for  $\varepsilon=0.02$  
(see Figs.~\ref{pre_fig1}(a) and ~\ref{pre_fig1}(b)),  while their time averaged 
frequencies are entrained (see the inset of Fig.~\ref{pre_fig1}(a)). Being identical R\"ossler oscillators
in the periodic regime, the frequencies of all the oscillators are always entrained
(see the insets of Fig.~\ref{pre_fig1}) in the entire parameter regimes we have traced.
The nonlocal coupling leads to the splitting of the ensemble into coexisting coherent and incoherent domains as depicted 
in Figs.~\ref{pre_fig1}(c) and ~\ref{pre_fig1}(d) for $\varepsilon=0.08$, thereby
confirming the existence of chimera state.  Further increase in the strength of
the nonlocal coupling results in the synchronous evolution of the ensemble of
R\"ossler oscillators.  Phase and complete synchronous evolution of 
the oscillators are clearly evident from  Figs.~\ref{pre_fig1}(e) and ~\ref{pre_fig1}(f),
respectively, for $\varepsilon=0.5$. Phase-flip chimera emerges after
the complete synchronized state, which is illustrated
in Figs.~\ref{pre_fig1}(g) and~\ref{pre_fig1}(h)  for  $\varepsilon=1.5$.
Phase-flip chimeras with two adjacent out-of-phase synchronized  
spatially coherent domains  interspersed by a spatially incoherent domain  
are clearly evident from the spatiotemporal plot in Fig.~\ref{pre_fig1}(h), while
their frequencies remain entrained. However, for the ease of visualization of the 
phase-flip chimera characterized by coexisting  out-of-phase synchronized coherent domains 
(where the nearby oscillators exhibit in-phase synchronized oscillations among them 
in each of the synchronized domains)
along with an asynchronous incoherent domain (where nearby oscillators exhibit
out-of-phase synchronized oscillations) interspersing the coherent domains at the 
phase-flip transition are enlarged and depicted in Fig.~\ref{pre_fig2}.
It is to be noted that the chimera states  investigated so far have emerged as an intermediate state 
in the transition from  completely incoherent to completely coherent states, in general.
In contrast, the phase-flip chimera emerges after the completely synchronized state.  Note that these are not steady states as in the case of chimera death states~\cite{zak2014,prema:15}, but evolve dynamically,  and they are also not two-cluster states as the oscillator index 
cannot be reordered.

\begin{figure*}
\centering
\includegraphics[width=2.0\columnwidth]{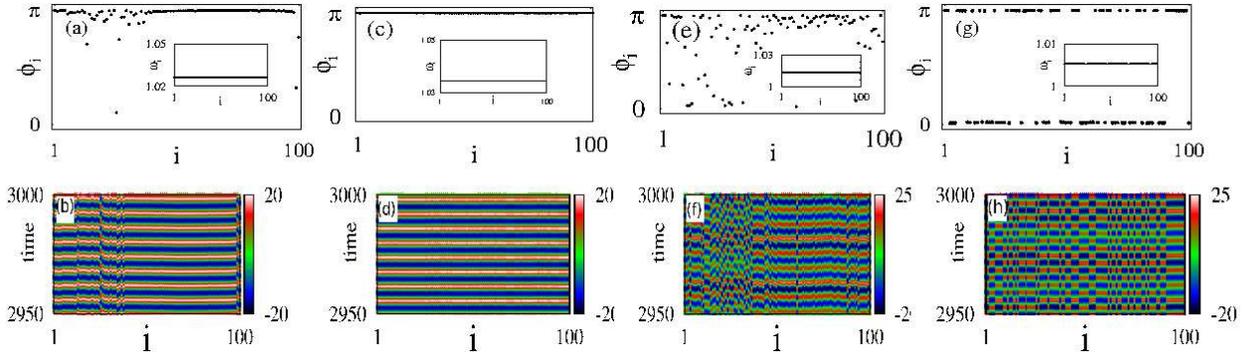}
\caption{(Color online) Snapshots of the instantaneous phases $\phi_{i}$ (top row) and the space-time evolution
 (bottom row) of the ensemble of R\"ossler oscillators in the chaotic regime for the coupling radius $r=0.3$ of the 
nonlocal coupling and for different values of the strength of the nonlocal coupling exhibiting 
(a)-(b) chimera state for $\varepsilon=0.5$,
(c)-(d) synchronized state for $\varepsilon=1.0$, 
(e)-(f)  desynchronized state for $\varepsilon=1.2$  and 
(g)-(h) phase-flip chimera for $\varepsilon=2.5$.
Time averaged frequencies of all the oscillators are shown in the insets. Other parameter values are $a=0.1$, $b=0.1$, $c=18$ and $\alpha=1$.}
\label{pre_fig3}
\end{figure*}
\section{Phase-flip chimera in an ensemble of R\"ossler oscillators in the chaotic regime}
For the parameters $a=0.1$, $b=0.1$ and $c=18.0$, the uncoupled R\"ossler oscillators
evolve in (\ref{ross}) chaotically.  Snapshots of the instantaneous
phases and the spatiotemporal evolution of the oscillators are depicted in
Fig.~\ref{pre_fig3} for different values of coupling, $\varepsilon$.  The time
averaged frequencies of all the oscillators are shown in the insets of Fig.~\ref{pre_fig3}. One can observe that  frequencies of the oscillators are entrained to the same frequency as indicated by a straight line of $w_i$.  For lower values of 
$\varepsilon$, the coupled oscillators evolve in asynchrony (not shown here), whereas the ensemble
of oscillators splits into coexisting coherent and incoherent domains, confirming
the existence of chimera, as shown in
Figs.~\ref{pre_fig3}(a) and ~\ref{pre_fig3}(b) for $\varepsilon=0.5$.
Increasing $\varepsilon$ further, the ensemble of oscillators evolve in 
complete synchrony (see Figs.~\ref{pre_fig3}(c) and ~\ref{pre_fig3}(d) for $\varepsilon=1.0$). 
The synchronized oscillators become desynchronized for further larger $\varepsilon$,
the dynamics of which is illustrated in the snapshot of the instantaneous phases and
the spatiotemporal plots in Figs.~\ref{pre_fig3}(e) and ~\ref{pre_fig3}(f), 
respectively, for $\varepsilon=1.2$. Phase-flip chimera emerges from the desynchronized
state upon increasing the strength of the nonlocal coupling further as depicted in
Figs.~\ref{pre_fig3}(g) and ~\ref{pre_fig3}(h) for $\varepsilon=2.5$. 

\begin{figure}
\centering
\includegraphics[width=1.0\columnwidth]{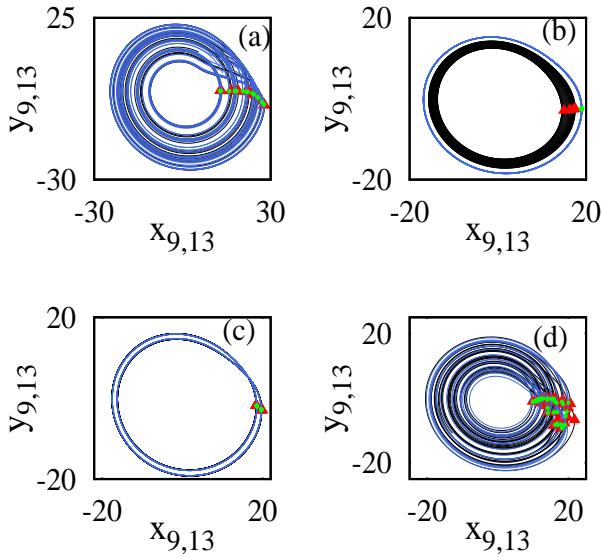}
\caption{(Color online) Representative class of attractors along with their Poincar\'e points 
throughout the dynamical transition of the ensemble of R\"ossler oscillators observed in 
Fig.~\ref{pre_fig3} as a function of the strength of the nonlocal coupling $\varepsilon$.
(a) desynchronized state for $\varepsilon=0.0$,
(b) chimera state for $\varepsilon=0.5$,
(c) synchronized state for $\varepsilon=1.0$,  and 
(d) desynchronized state for $\varepsilon=1.2$. In Fig (a) the value of $k=0$, while in (c)-(d) it is fixed as 10.}
\label{pre_fig4}
\end{figure}
The dynamics of the individual R\"ossler oscillators changes and plays a crucial
role in the emergence of the observed collective dynamical behaviors of the ensemble of R\"ossler oscillators
in Fig.~\ref{pre_fig3} as a function of $\varepsilon$. 
The R\"ossler oscillators always exhibit periodic oscillations throughout the entire dynamical
transition regimes of the ensemble discussed in Fig.~\ref{pre_fig1}, whereas
in the case of chaotic oscillations of the individual oscillators the dynamical nature of
the oscillators changes as will be discussed below. In the absence of the couplings $k=0$
and $\varepsilon=0$, the uncoupled individual R\"ossler oscillators exhibits chaotic oscillations
as pointed above.  A couple of uncoupled representative R\"ossler oscillators (namely $N=9$ and $13$)  exhibiting chaotic
oscillations are depicted in Fig.~\ref{pre_fig4}(a).  The filled circles and triangles in Fig.~\ref{pre_fig4}
are the Poincar\'e points. The value of $k$ in the other figures ~\ref{pre_fig4}(b)-(d) is fixed as $k=10$ as  in Fig.~\ref{pre_fig3}. The ensemble of R\"ossler oscillators  displays chimera state for $\varepsilon=0.5$
(see Fig.~\ref{pre_fig3}(a)). A representative oscillator from each of the coherent and incoherent domains
is displayed in Fig.~\ref{pre_fig4}(b). The synchronized oscillators 
in the coherent domain are entrained to periodic oscillations as indicated by the blue (light grey) line in 
Fig.~\ref{pre_fig4}(b), while the asynchronous oscillators in the incoherent domain exhibit
chaotic oscillations as indicated by the black (dark grey) line in Fig.~\ref{pre_fig4}(b). The ensemble
of oscillators are synchronized for the strength of the nonlocal coupling $\varepsilon=1.0$ (see Fig.~\ref{pre_fig3}(c)),
where the dynamics of the ensemble of oscillators become periodic as indicated by the 
representative oscillators in Fig.~\ref{pre_fig4}(c). The ensemble of  R\"ossler oscillators are
desynchronized as already shown in Fig.~\ref{pre_fig3}(e) for  $\varepsilon=1.2$ rendering the
oscillators to exhibit chaotic oscillations as displayed in Fig.~\ref{pre_fig4}(d) and the dynamics of the
individual  R\"ossler oscillators  remain chaotic for further larger values of $\varepsilon$.

\begin{figure}
\centering
\includegraphics[width=1.0\columnwidth]{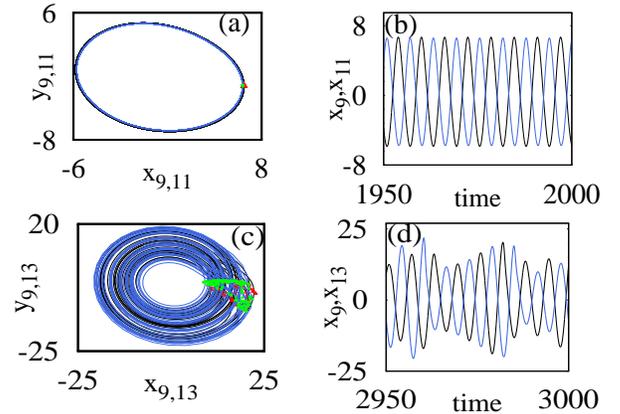}
\caption{(Color online) Representative class of attractors along with their Poincar\'e points 
and their time evolution plots
of the out-of-phase synchronized coherent domains of the phase-flip chimera observed in 
Figs.~\ref{pre_fig1}(g) and~\ref{pre_fig3}(g).  Top row represents the R\"ossler oscillators in the periodic regime
and bottom row corresponds to the R\"ossler oscillators exhibiting chaotic oscillations.}
\label{pre_fig5}
\end{figure}
Attractors and the corresponding time series plots of the representative oscillators from the out-of-phase
synchronized coherent domains of the phase-flip chimera are depicted in 
Fig.~\ref{pre_fig5}. The  top row represents the R\"ossler oscillators in the periodic regime
and the bottom row corresponds to the R\"ossler oscillators exhibiting chaotic oscillations.  
Attractors exhibting both in-phase and anti-phase oscillations for the periodic case
and chaotic case along with their Poincar\'e points are shown in Figs.~\ref{pre_fig5}(a) and \ref{pre_fig5}(c), respectively. 
Time series plots in  Figs.~\ref{pre_fig5}(b) and \ref{pre_fig5}(d) clearly displays the out-of-phase oscillations of
the two adjacent coherent domains of the phase-flip chimera.

\section{\label{quanti}Quantification measures to characterize the chimera states}
The notion of the strength of incoherence $S$~\cite{Gopal1:14}, was recently introduced by Gopal et al to characterize and
to distinguish various collective dynamical states, defined as
\begin{equation} 
S=1-\frac{\sum_{m=1}^{M}s_{m}}{M}, \hspace{0.1cm}  s_{m}=\Theta(\delta-\sigma_{l}(m)),
\end{equation}
where $\Theta(\cdot)$ is the Heaviside step function, and $\delta$ is a predefined 
threshold. Normally, $\delta$ is chosen as a certain percentage value of the difference between
the upper/lower bounds, $x_{l,i_{max}}/x_{l,i_{min}}$,  of the allowed values of $x_{l,i}$.
$M$ is the number of bins of equal size $n=N/M$.
The local standard deviation $\sigma_{l}(m)$ is introduced as
\begin{equation}
\sigma_{l}(m)=\Big<\noindent \sqrt{\frac{1}{n}\sum_{j=n(m-1)+1}^{mn}[z_{l,j}-<z_{l,m}>]^2} 
\hspace{0.1cm}\Big>_{t}, \hspace{0.1cm}m=1,2,...M,
\end{equation}
where $z_{l,i}=x_{l,i}-x_{l,i+1}$, $l=1,2...d$, $d$ is the dimension of the individual unit in
the ensemble, $i=1,2...N$, $<z_{l,m}>=\frac{1}{n}\sum_{j=n(m-1)+1}^{mn}z_{l,j}(t)$, and
$\langle ...\rangle_{t}$ denotes the time average.
When $\sigma_{l}(m)$ is less than $\delta$, $s_{m}=1$, otherwise $s_{m}=0$ ($m$ in the present case is chosen as $m=20$). 
The local standard deviation $\sigma_{l}(m)$ has 
some finite value in the incoherent domain $\forall~m$, which is always greater than 
$\delta$ and hence  $s_{m}=0, \forall~m$, thereby resulting in unit value for the
strength of incoherence $S$  in the incoherent domain. On the other hand, the
standard deviation $\sigma_{l}(m)$ is always zero in the coherent domain and 
hence $s_{m}=1, \forall~m$, thereby resulting in the null value of $S$.
Since the chimera states are characterized by coexisting coherent and incoherent domains,
the strength of incoherence $S$ will have intermediate values between zero and one, $0 < S < 1$. 

\begin{figure}
\centering
\includegraphics[width=1.0\columnwidth]{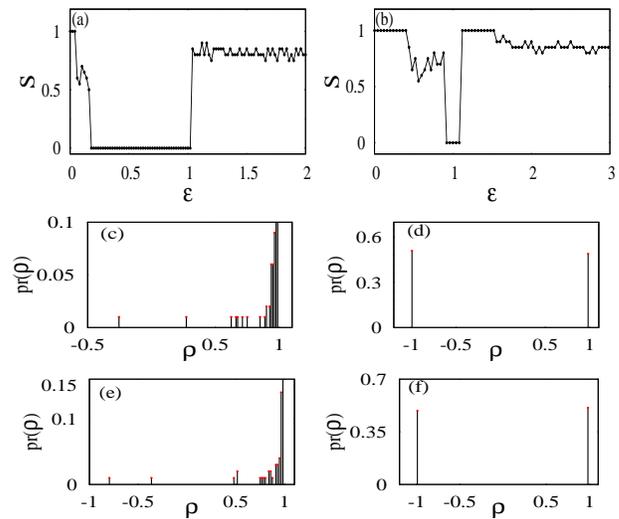}
\caption{(Color online) The strength of incoherence $S$ as a function of the strength of
the nonlocal coupling $\varepsilon$ in (a) periodic case, (b) chaotic case. The probability 
distribution pr$(\rho)$ of the correlation coefficient (c)  conventional chimera in the periodic case (d)  phase-flip chimera in the periodic case, (e)   conventional chimera in the chaotic case 
and (f)  phase-flip chimera in the chaotic case.}
\label{pre_fig6}
\end{figure}
The strength of incoherence is shown in Figs.~\ref{pre_fig6}(a) and~\ref{pre_fig6}(b)
as a function of the strength of the nonlocal coupling $\varepsilon$
characterizing the dynamical transition discussed in Figs.~\ref{pre_fig1} and~\ref{pre_fig3},
respectively. Unit value of $S$ in the range of $\varepsilon\in(0,0.08)$ in Fig.~\ref{pre_fig6}(a) corroborates the
asynchronous evolution of the R\"ossler oscillators with periodic oscillations.  Intermediate value of $S$ between  zero
and unity confirms the existence of conventional chimera in the range of $\varepsilon\in(0.08,0.18)$.
Null value of $S$  in the range of $\varepsilon\in(0.18,1.02)$ attributes to the synchronous evolution
of the ensemble of R\"ossler oscillators. Fluctuations in the value of $S$ close to unity
elucidates the existence of phase-flip chimera state. As the phase-flip chimera is characterized
by out-of-phase synchronized coherent domains interspersed by an incoherent domain,
most of the bins during this state have a mixture of in-phase and anti-phase oscillations, while
a few bins may have completely coherent (either in-phase or anti-phase) oscillations  leading to
fluctuating values of $S$ close to unity.
Now, we will discuss the dynamical transtion observed in Fig.~\ref{pre_fig3} for the chaotic oscillations
of the individual R\"ossler oscillators in terms of 
the strength of incoherence $S$ displayed in Fig.~\ref{pre_fig6}(b). Asynchronous evolution
of the oscillators are indicated by the unit value of $S$ in the range of $\varepsilon\in(0,0.48)$,
whereas the intermediate value of $0 < S < 1$ confirms the existence of conventional chimera
in the range of $\varepsilon\in(0.48,0.96)$. Narrow range of synchronized state
is confirmed by $S=0$ in the range of $\varepsilon\in(0.96,1.04)$. The oscillators get
desynchronized in the range of $\varepsilon\in(1.04,1.66)$ as indicated by a unit
value of $S$. Fluctuations in $S$ close to unity beyond $\varepsilon=1.66$ corroborates
the existence of phase-flip chimera.

We have also estimated the probability distribution of the correlation coefficient defined as
\begin{equation}
\rho_i = \frac{\langle(x_{1}(t)-\langle x_{1}(t)\rangle)(x_{i}(t+\Delta t)-\langle x_{i}(t)\rangle)\rangle}{\sqrt{\langle(x_{1}(t)-\langle x_{1}(t)\rangle)^{2}\rangle_t \langle(x_{i}(t)-\langle x_{i}(t)\rangle)^{2}\rangle}_t},
\label{prob}
\end{equation}
where i=1,2..,N, to characterize the conventional chimera and phase-flip chimera. In Eq.~(\ref{prob}), $\langle\cdot \rangle_t$ represents the time average and  $\Delta t$ is the time shift. The correlation coefficient is estimated by using each oscillator in the ensemble
as  a reference oscillator  and averaging it over the number of oscillators $N$.
The probability distribution of the correlation coefficient is depicted in 
Figs.~\ref{pre_fig6}(c)-\ref{pre_fig6}(f) for different values of $\varepsilon$ in both the 
periodic (Figs.~\ref{pre_fig6}(c) and \ref{pre_fig6}(d)) and chaotic (Figs.~\ref{pre_fig6}(e) and \ref{pre_fig6}(f)) regimes.
Correlation coefficient for completely synchronized oscillators acquire unit value and for
antisynchronous state acquires $-1$, whereas for the desynchronous state it is characterized by
the intermediate values between $\pm 1$.  
Since the conventional chimera is characterized by the coexistence of synchronized 
and asynchronous domains, the probability distribution of the correlation coefficient
is large near unit value and small at other values of correlation coefficient as 
seen in Figs.~\ref{pre_fig6}(c) and \ref{pre_fig6}(e).  On the other hand, the phase-flip chimera
is characterized by only two inhomogeneous states, namely in-phase and anti-phase synchronized
states, the correlation coefficient of the phase-flip chimera acquires only $+1$ and $-1$ 
as its values.  Consequently, the probability distribution of the correlation coefficient
at phase-flip chimera has only two values at $+1$ and $-1$ as evident from  Figs.~\ref{pre_fig6}(d) and \ref{pre_fig6}(f).

\begin{figure}
\centering
\includegraphics[width=1.00\columnwidth]{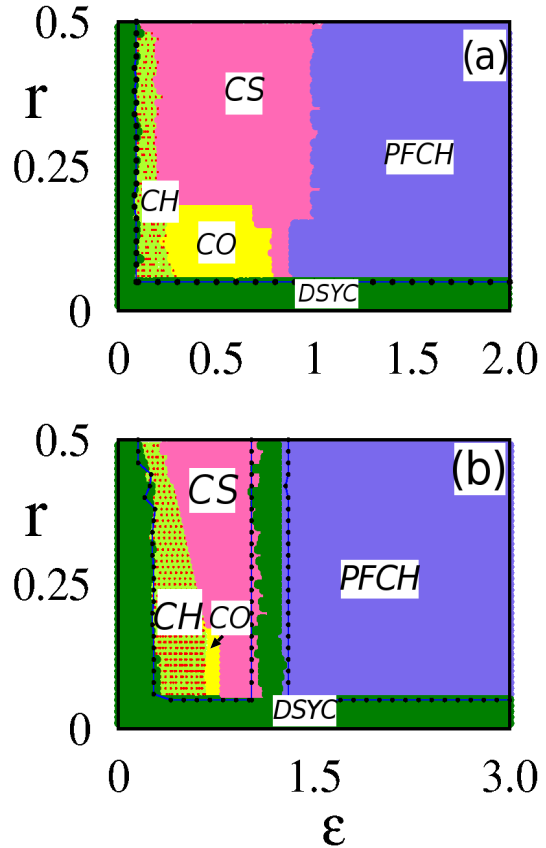}
\caption{(Color online) Two parameter phase diagram depicting the collective dynamical
states of the ensemble of R\"ossler oscillators with common dynamic environment as
a function of the strength of the nonlocal coupling $\varepsilon$ and 
the coupling radius $r$. (a) Periodic state (b) Chaotic state. The parameter spaces marked as
$DSYC$, $CH$, $PFCH$, $CO$, and $CS$ correspond to the desynchronized state,
conventional chimera, phase-flip chimera, coherent state and
complete synchronized state, respectively. The dotted lines correspond to
the stability curves estimated from the eigenvalues of the variational equation (\ref{ross2}).}
\label{pre_fig7}
\end{figure}
\section{Two-parameter phase diagrams}
A two-parameter phase diagram as a function of the strength of the nonlocal coupling
$\varepsilon\in(0,2)$ and the coupling radius $r\in(0,0.5)$ is depicted in 
Fig.~\ref{pre_fig7}(a) to gain a global perspective of collective dynamics emerging from
the ensemble of R\"ossler oscillators exhibiting periodic oscillations in a 
dynamic environment with nonlocal coupling.  We have used the above quantification measures
discused in Sec.~\ref{quanti} to demarcate the different dynamical regimes of the ensemble of
R\"ossler oscillators. The dynamical regimes delineated 
in Fig.~\ref{pre_fig1} are indicated as $DSYC$, $CH$, $SYC$ and $PFCH$, which
correspond to the desynchronized states, chimera states, synchronized states
and phase-flip chimeras, respectively.  In addition
to the above regimes, we have also found the coherent states marked as $CO$,
where all the oscillators evolve in coherence  in the range of the coupling radius
$r\in(0.05,0.2)$ and  $\varepsilon\in(0.2,0.8)$.

Using the well known master stability function (MSF) formalism~\cite{Pecora:98,asrea2015}, the entire parameter
space in Fig.~\ref{pre_fig7} can be demarcated into desynchronized state ($DSYC$)
and synchronous state ($SYC$).
The stability of the synchronized manifold $(x_i=x, y_i=y, z_i=z$ and $w_i=w$; $\forall~i$)  
is determined by the variational equations 
\begin{subequations}
\begin{align}
\dot{\eta}_{1j}=&\,-{\eta}_{2j}-{\eta}_{3j},  \\ 
\dot{\eta}_{2j}=&\,{\eta}_{1j}+a{\eta}_{2j}, \\ 
\dot{\eta}_{3j}=&\,{\eta}_{3j}(x(t)-c)+k{\eta}_{4j}+z(t){\eta}_{1j}, \\  
\dot{\eta}_{4j}=&\,-\alpha{\eta}_{4j}+\frac{{\eta}_{3j}}{2}+\varepsilon \lambda_j {\eta}_{4j}, 
\end{align}
\label{ross1}
\end{subequations}
where $x(t)$ and $z(t)$ are the solution of the uncoupled equation (\ref{ross}). 
$\eta_{ik}=\bf{\zeta}_i \bf{Q}_k$, where $\bf{\zeta}_i=(\zeta_{i1},\zeta_{i2},\ldots, \zeta_{iN})$, and $(\bf{\zeta}_1,\bf{\zeta}_2,\bf{\zeta}_3,\bf{\zeta}_4)$ are the deviation of $(\bf{x},\bf{y},\bf{z},\bf{w})$ from the synchronized solution $(x,y,z,w)$.  $\bf{Q}_k$ is the eigenvector of the coupling matrix $G$ and $\lambda_j$  are the eigenvalues, where 
\begin{equation}
G=\begin{bmatrix}
    a_{11} & a_{12} & a_{13} & \dots  & a_{1n} \\
    a_{21} & a_{22} & a_{23} & \dots  & a_{2n} \\
    \vdots & \vdots & \vdots & \ddots & \vdots \\
    a_{d1} & a_{d2} & a_{d3} & \dots  & a_{dn},
\end{bmatrix}
\end{equation}
with $a_{ii}=-1$ and $a_{i(i+j)}=a_{i(i-j)}=\frac{1}{2P}$ for $i=1,2,...,N$ and $j=1,2,... P$. 
Here  $a_{i(N+k)}=a_{ik}$ and $a_{i(1-k)}=a_{i(N-k+1)}$  for $k=1,2,... P$. The eigenvalues are given by 
\begin{align}
\lambda_j =-1+\frac{1}{P}\sum_{k=1}^{P}\cos\left(\frac{2\pi k}{N}j\right), \;\; j=0,1,2,\ldots,N-1.
\end{align}
The eigenvalue $\lambda_0$ corresponds to the perturbation parallel to the synchronization
manifold, while the other $N-1$ eigenvalues correspond  to the perturbation transverse
to the synchronization manifold.  The transverse eigenmodes should be damped
out to have a stable synchronization manifold. The stability of the synchronization
manifold depends only on the largest eigenvalue 
$\lambda_1 =-1+\frac{1}{P}\sum_{k=1}^{P}\cos(\frac{2\pi k}{N})$
and the corresponding variational equation becomes
\begin{align}
\dot{\eta}_{11}=&\,-{\eta}_{21}-{\eta}_{31},   \nonumber  \\ 
\dot{\eta}_{21}=&\,{\eta}_{11}+a{\eta}_{21},  \nonumber  \\   
\dot{\eta}_{31}=&\,{\eta}_{31}(x(t)-c)+k{\eta}_{41}+z(t){\eta}_{11}, \nonumber   \\ 
\dot{\eta}_{41}=&\,-\alpha{\eta}_{41}+\frac{{\eta}_{31}}{2}-\varepsilon (1-\frac{1}{P}\sum_{k=1}^{P}\cos(\frac{2\pi k}{N})) {\eta}_{41}. 
\label{ross2}
\end{align}

\begin{figure}
\centering
\includegraphics[width=0.9\columnwidth]{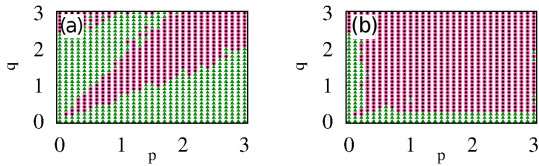}
\caption{(Color online) Basin of attraction depicting the coexisting phase-flip chimera (red, $\bullet$) and complete synchronized states (green, $\blacktriangle$) in (a) periodic and (b) chaotic cases of the coupled R\"ossler oscillators.}
\label{pre_fig8}
\end{figure}

Now, the value of the largest Lyapunov exponent of the variational equation
(\ref{ross2}) for each value of the parameters in the two-parameter phase diagram (Fig.~\ref{pre_fig7})
is used to demarcate the two-parameter phase diagram into desynchronized state ($DSYC$) and synchronous state ($SYC$)
as indicated by the dotted lines in Fig.~\ref{pre_fig7}(a). Before the dotted lines the
largest Lyapunov exponent acquires positive values while for the parameters  above the dotted lines
it acquires negative values attributing to the stability of the synchronization manifold.  Figure~\ref{pre_fig7} elucidates that
the simulation results are in agreement with the results obtained using the semi-analytic approach,
namely the master stability function formalism in demarcating the synchronized and desynchronized regimes.
 It is to be noted that the synchronous state is a multistable state with coexisting chimera states, coherent states
and phase-flip chimeras depending on the distribution of the initial conditions (for a given choice of $\epsilon$ and $r$ or $\epsilon$ and $k$).  The multistability
nature of the synchronized parameter space coexisting along with the phase-flip chimera will be
discussed in Fig.~\ref{pre_fig8}.

A similar two-parameter phase diagram in the range of $\varepsilon\in(0,3)$ and $r\in(0,0.5]$
of the ensemble of R\"ossler oscillators exhibiting chaotic oscillations is
shown in Fig.~\ref{pre_fig7}(b).  The main difference between the emergent
dynamics from the ensemble of R\"ossler oscillators with periodic oscillations
(see Fig.~\ref{pre_fig7}(a)) and that with chaotic oscillations (see Fig.~\ref{pre_fig7}(b)) is that  in the later case the 
 phase-flip chimera (marked as $PFCH$) is immediately
preceded by the asynchronous oscillations, indicated by $DSYC$,  of the ensemble.
That is, the dynamical transition occurs in the sequence of desynchronized state ($DSYC$),
chimera state ($CH$), coherent state ($CO$) or completely synchronized
state ($SYC$) depending on the value of $r$,  again desynchronized state ($DSYC$) 
and finally  phase-flip chimera ($PFCH$)
as a function of $\varepsilon$, which is evident from the two-parameter phase diagram 
in  Fig.~\ref{pre_fig7}(b). The synchronized and desynchronized parameter 
space is also demarcated  by solving the variational equation Eq.~(\ref{ross2}) 
as indicated by the dotted lines in Fig.~\ref{pre_fig7}. Here, the parameter space enclosed
between the dotted lines are characterized by the negative values of the largest Lyapunov exponent
of the variational equations attributing to the stable synchronized state.
It is to be noted that in both the two-parameter phase diagrams,
the discussed transition occurs for $r>0.05$, below which the ensemble of oscillators
remain in asynchronous state elucidating that the phase-flip chimera may not emerge in  the
nearest neighbor coupling between the agents. It also reveals that an appropriate
coupling radius is necessary for the emergence of phase-flip chimera. 

In order to demonstrate the stability of the phase-flip chimera, we have depicted
the basin of attraction of the ensemble, Eq.~(\ref{ross}), in Fig.~\ref{pre_fig8}
as a function of $p$ and $q$, which correspond to the initial conditions of $x$ and $y$ variables.  
The initial conditions for the variables $z$ and $w$
are uniform random numbers distributed between $-1$ and $1$, whereas that of
$x$ and $y$ variables are the uniform random numbers distributed between 
$\pm p$ and $\pm q$,  respectively.  It is clear that both the phase-flip chimeras 
(indicated by circles) and completely synchronized states (indicated by triangles) coexist  for a wide choice of
initial conditions corroborating the stability and robustness of the phase-flip chimera.

\begin{figure}
\centering
\includegraphics[width=0.9\columnwidth]{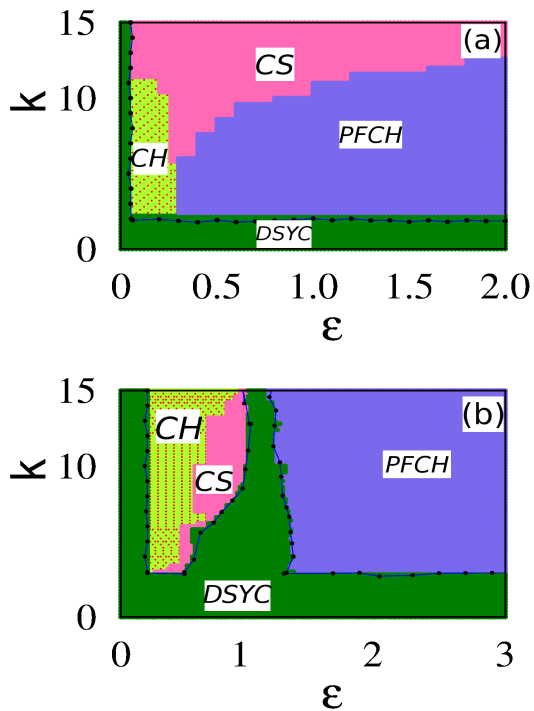}
\caption{(Color online) Two parameter phase diagram depicting the collective dynamical states of the ensemble of 
 R\"ossler
oscillators with common dynamic environment as a function of $k$ and $\varepsilon$. The parameter
space marked as $DSYC, CH, PFCH$ and $CS$ corresponds to the desynchronized state, conventional chimera, 
phase-flip chimera, and complete synchronized state, respectively. The dotted lines corresponds to
the stability curves estimated from the eigenvalues of the variational equation (\ref{ross2}).}
\label{pre_fig9}
\end{figure}
To explore the effect of the environmental coupling on the observed dynamical transition 
in detail, we have plotted the two-parameter phase diagram as a function of the strength
of the nonlocal coupling $\varepsilon$ and the strength $k$ with which the local agents
interact with their respective oscillators in Figs.~\ref{pre_fig9}(a)
and~\ref{pre_fig9}(b) for periodic and chaotic oscillations of the individual uncoupled
R\"ossler oscillators, respectively. The parameter space leading to desynchronous
state, conventional chimera, phase-flip chimera and complete synchronous state
are marked as $DSYC, CH, PFCH$ and $CS$, respectivey, in both the figures. 
Asynchronous and completely synchronous parameter space are also demaracted (indicated by
dotted lines) using the value of the largest Lyapunov exponent of the variational equation (\ref{ross2}).
In the periodic regime of the R\"ossler oscillators,  the coupled oscillators remain asynchronous 
in the entire range of $\varepsilon$ for small values of $k$ as seen in Fig.~\ref{pre_fig9}(a).
In the range of $k\in(2.1,5.6)$, desynchronized state is followed by conventional chimera and
the phase-flip chimera as  $\varepsilon$ is increased. Increasing $k$ further, the conventional chimera
and phase-flip chimera regimes are seperated by complete synchronized regime upto to $k=12.6$.
Transition from asynchronous to complete synchronized state occurs for $k>12.6$ as a function
of $\varepsilon$. Thus the emergence of phase-flip chimera is spread over a wide range of $k\in(2.1,12.6)$
and $\varepsilon$. For the chaotic oscillations of the individual R\"ossler oscillators,  we have 
observed similar transitions as in Fig.~\ref{pre_fig9}(a) as a function of $k$ and $\varepsilon$ except
for the fact that the range of phase-flip chimera extends to much wider range of $k$ and  $\varepsilon$
(See Fig.~\ref{pre_fig9}(b)) while the desynchronized state preceeds the phase-flip chimera. 

The existence of phase-flip bifurcation/transition in two coupled oscillators induced by the environmental coupling in Eq.(\ref{ross}) was shown in~\cite{Ambi:10,Amit:12}. We find that the environmental coupling among the agents and the oscillators induces phase-flip transition/bifurcation while the nonlocal coupling among the agents in the common environment facilitates the onset of incoherent domain at the phase-flip transition. In the absence (or for low values of the strength) of the environmental coupling one cannot observe the phase-flip chimera as is evident from the two-phase diagram in Fig.~\ref{pre_fig9}. We have also found that the emergence of the phase-flip chimera depends on the relaxation time of the environmental coupling in Eq.~1(d) and phase-flip chimera emerges  for the parameter $\alpha\geq1$, which determines the relaxation time of the external agent $w_{i}$  (See Appendix A for more details).

\begin{figure}
\centering
\includegraphics[width=1.0\columnwidth]{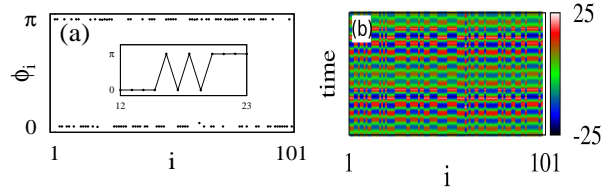}
\caption{(Color online) (a) Snapshots of the instantaneous phases $\phi_{i}$ and (b) the space-time evolution
of the ensemble of R\"ossler oscillators in the chaotic regime with global coupling between the agents for the
value of the strength of the global coupling $\varepsilon=2.5$.}
\label{pre_fig10}
\end{figure}
\section{Phase-flip chimera with globally coupled agents}
 It is also to be noted that we have identified the emergence of 
the phase-flip chimera in the case of globally coupled agents.
Snapshots of the instantaneous phases $\phi_{i}$ and the space-time evolution
of the ensemble of R\"ossler oscillators in the chaotic regime but with global coupling between the agents for the
value of the strength of the global coupling $\varepsilon=2.5$ is shown in Fig.~\ref{pre_fig10}. The value of the other
parameters and the distribution of the initial conditions are the same as in Fig.~\ref{pre_fig3}. 
It is clear from the figure that the phase-flip chimera characterized by two adjacent out-of-phase
synchronized coherent domains, where nearby oscillators are exhibiting in-phase synchronized oscillations,
interspersed by an incoherent domain, where the nearby oscillators exhibit out-of-phase oscillations
(see the inset of Fig.~\ref{pre_fig10}), exists even 
with the global coupling between the agents in the common environment.  We have also
confirmed the emergence of phase-flip chimera from the ensemble for the periodic oscillations of the
individual R\"ossler oscillators with the global coupling between the agents.
It is known that in an ensemble of globally coupled oscillators one can reorder the spatial
index of the oscillators such that the spatial inhomogenity with coexisting in-phase 
and out-of-phase synchronized oscillators in the phase-flip chimera 
state (see Fig.~\ref{pre_fig2}) can be recasted as a two-cluster 
state~\cite{manrubia:04,okuda:93,zanette:98}. 
However, we would like to point out that once the spatial ordering of the globally coupled 
oscillators are fixed by indexing them from $1$ to $N$, then the distribution of initial 
conditions as stated above among the oscillators results in spatially inhomogeneous
states with coherent out-of-phase synchronized domains interspersed by incoherent domain
comprised of nearby oscillators exhibiting in-phase and anti-phase oscillations.
This confirms the emergence of phase-flip chimera even with global coupling among the agents. Indeed, several recent investigations reported the existence of chimera states in an
ensemble of globally coupled oscillators~\cite{Sethia1:14,baner:14,Gopal1:14,sch:15,sch1:15,bolo:16}. 

Finally, we have also checked whether phase-flip chimeras occur when the coupling is additionally  given to $x$ or $y$ variables that is to Eq.~(\ref{ross}a) or Eq.~(\ref{ross}b). In these cases only transition from desynchronized state to synchronized state occurs (see Appendix B for more details).

\section{Conclusions}
In summary, we have identified an interesting type of a collective dynamical regime,
called phase-flip chimeras, in an ensemble of identical R\"ossler oscillators coupled 
indirectly via the agents from  a common dynamic environment, where the agents are coupled nonlocally with
a coupling radius $r$.  Such interactions are found in the diffusion 
of biomolecues between the cells and their environment~\cite{Ullner:07}.
The phase-flip chimera is characterized by two out-of-phase synchronized 
spatially coherent domains interspersed by a spatially incoherent domain comprised of nearby oscillators 
exhibiting out-of-phase oscillations. The oscillators
in  each of the coherent domains exhibit phase synchronized oscillations among themselves, whereas 
the two adjacent coherent domains in the phase-flip chimera exhibit out-of-phase synchronized oscillations.
The robustness of the phase-flip chimera is also confirmed by depicting its occurrence 
in a wide range of parameters using the two-parameter phase diagram. Further, it is also shown that
the phase-flip chimera is preceded by the conventional chimera and it
emerges only after the completely synchronized state emerges.
In the chaotic regime the phase-flip chimera is immediately 
preceded by asynchronous state, which in turn emerges as a desynchronized state
from the synchronous evolution of the ensemble of oscillators. We have used the strength of incoherence,
probability distribution of the correlation coeffecient and the master stability function to characterize
the observed dynamical transition of the ensemble of the  R\"ossler oscillators. It is also confirmed
that the phase-flip chimera emerges even in the global coupling among the agents, whereas
with the nearest neighbor coupling such a state does not emerge.

\section*{acknowledgments}
The work of VKC forms part of a research project sponsored by INSA Young Scientist Project under Grant No. SP/YSP/96/2014. DVS is supported by the SERB-DST Fast Track scheme for young
scientist under Grant No. ST/FTP/PS-119/2013. ML  is supported by a NASI Platinum Jubilee Senior Scientist Fellowship.

\begin{appendix}
\section*{Appendix A: Relaxation time of the environmental coupling Eq.~(\ref{ross4})}

\begin{figure}
\centering
\includegraphics[width=0.8\columnwidth]{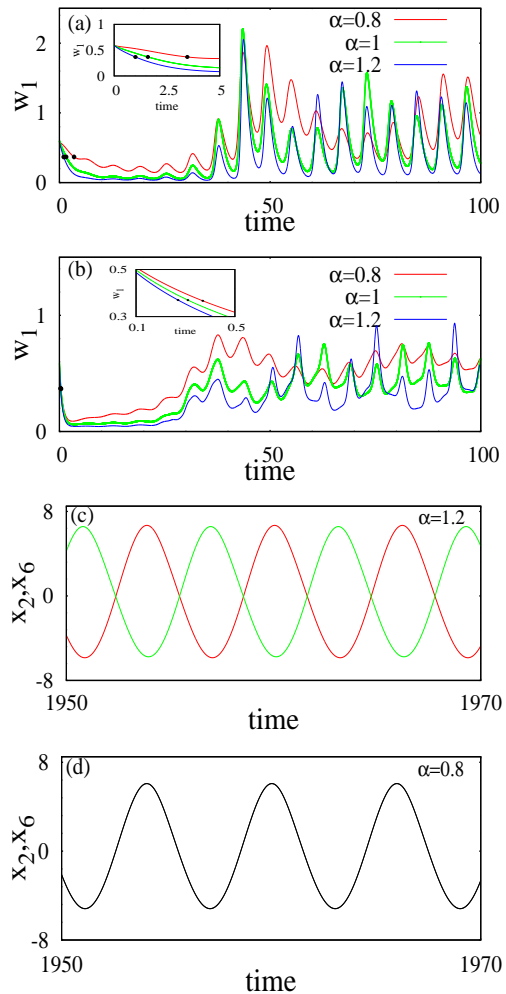}
\caption{(Color online) Dynamics of the agent $w_{1}$ is shown in (a) for $\varepsilon=0$, and
(b) for $\varepsilon=1.5$. In both the figures $k=10$. The dynamics of random oscillators are shown in (c) and (d) for different values of the relaxation parameter $\alpha$.}
\label{time}
\end{figure}

The dependence of the dynamics of the oscillators on the relaxation time is depicted in  
Fig.~\ref{time}.  Dynamics of the agent $w_{1}$ (as an example) is shown in Fig. \ref{time}(a) for $\varepsilon=0$, and in Fig. \ref{time}(b) for $\varepsilon=1.5$.  The dynamics of randomly chosen oscillators are shown in Figs. \ref{time}(c) and \ref{time}(d) for different values of the relaxation parameter $\alpha$ occurring in Eq.~(\ref{ross4}).

As the value of the parameter $\alpha$ is increased the relaxation time (pointed out by 
filled circles in the insets) decreases both in the absence (Fig.~\ref{time}(a)) and in 
the   presence (Fig.~\ref{time}(b)) of the coupling between the agents $w_i$ in 
Eq.~(\ref{ross4}).  It is to be noted that the relaxation time is very less in the 
presence of the coupling between the oscillators (See Fig.~\ref{time}(b)) compared 
to the uncoupled oscillators (See Fig.~\ref{time}(a)). Further, the value of the 
parameter $\alpha$ which determines the relaxation time dictates the collective behavior 
of the ensemble of the oscillators.  Phase-flip chimeras are observed for $\alpha\geq1$. 
For $\alpha=1$ the phase-flip chimera is already shown in Figs. 1 and 3. For $\alpha>1.2$, 
the out-of-phase oscillations of the adjacent out-of-phase synchronized coherent domains 
of the phase-flip chimera are represented by a couple of representative oscillators in 
Fig.~\ref{time}(c).  On the other hand, for $\alpha<1.0$, we observe only synchronized 
state for any choice of initial conditions and coupling strength as shown by the representative 
oscillators in Fig.~\ref{time}(d).

Thus we find that the emergence of the phase-flip chimera depends on the relaxation time of 
the environmental coupling in Eq.~(\ref{ross4}) and that it emerges  for the parameter 
$\alpha\geq 1$, which determines the relaxation time of the $w_i$'s. 

\begin{figure}
\centering
\includegraphics[width=1.0\columnwidth]{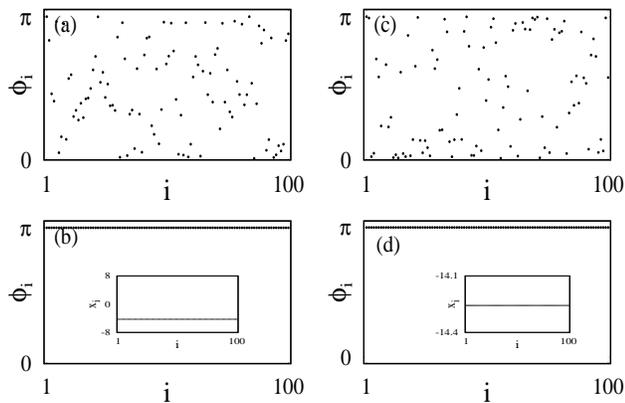}
\caption{(Color online) Snapshots of the instantaneous phases $\phi_{i}$ of the ensemble of R\"ossler oscillators
coupled via $y$-variable displaying (a) desynchronized state for $\varepsilon=0.005$, (b)
completely synchronized state for $\varepsilon=0.02$, (c) desynchronized state for $\varepsilon=0.005$  and (d)
completely synchronized state for  $\varepsilon=0.04$. Left column for periodic oscillations and right column for
chaotic oscillations of the R\"ossler oscillators. Insets in Fig.(b) and (d) depict the snapshots of $x_i$ .}
\label{otherv}
\end{figure}

\section*{Appendix B: Dynamics of coupling with other variables}
We have also examined the typical transition from desynchronized state to synchronized states 
by coupling the agent $w_{i}$ to the other variables $x_{i}$ or $y_{i}$ in the same fashion as in Eq.~(\ref{ross4}). 
Snapshots of the instantaneous phases $\phi_{i}$ of the ensemble of R\"ossler oscillators
coupled via the $y$-variable are shown in Fig.~\ref{otherv}. The left column corresponds to
the case of periodic oscillations when the oscillators are uncoupled, while the right column corresponds to the case of the chaotic oscillations of the uncoupled oscillators. Desynchronized state is observed for $\varepsilon=0.005$ (See Fig.~\ref{otherv}(a)) for the case of periodic oscillations.  Upon increasing the coupling
strength the oscillators are completely synchronized as shown in Fig.~\ref{otherv}(b)
for $\varepsilon=0.02$ and remains synchronized for any large coupling strength. 
For the case of chaotic oscillations of the uncoupled R\"ossler oscillators,
the oscillators are desynchronized for low values of the coupling strength as
shown in Fig.~\ref{otherv}(c) for $\varepsilon=0.005$ and remain completely
synchronized above a threshold value of $\varepsilon$ as illustrated in 
Fig.~\ref{otherv}(d) for $\varepsilon=0.04$. Hence, only the existence
of completely synchronized state from the desynchronized state is found and
we are not able to find phase-flip chimera for any value of $k$ and $\varepsilon$. Similar scenario arises for $x$ coupling as well (which we do not present here).

\end{appendix}


\begin{thebibliography}{39}
\bibitem{Kuramoto:84}
Y. Kuramoto, {\em Chemical Oscillations, Waves, and Turbulence}, (Springer-Verlag., Berlin, 1984).

\bibitem{Pikovsky:01}
A. Pikovsky, M. Rosenblum, and J. Kurths, {\em Synchronization —A Universal Concept	in Nonlinear Sciences} (Cambridge University Press, Cambridge, England, 2001).

\bibitem{Winfree:01}
A. T. Winfree, {\em The Geometry of Biological Time} (Springer, New York, 2001).

\bibitem{Kuramoto:1}
D.~Battogtokh and Y.~Kuramoto, Nonlinear. Phenom. Complex. Sys. \textbf{5},380 (2002);
 D. M. Abrams and S. H. Strogatz Phys. Rev. Lett. {\bf93}, 174102 (2004); D. M. Abrams, R. Mirollo, S. H. Strogatz and D. A. Wiley, Phys. Rev. Lett. {\bf101}, 084103 (2008); G. C. Sethia, A. Sen and F. M. Atay, Phys. Rev. Lett. \textbf{100}, 144102 (2008); M. J. Panaggio and D. M. Abrams, Nonlinearity 28, R67 (2015).

\bibitem{laing2000}
C.~.R.~Laing and C.~C.~Chow, Neural Computation {\bf 13}, 1473 (2000); C. R. Laing, Phys. Rev. E 81, 066221 (2010);
G. Bordyugov, A. Pikovsky, and M. Rosenblum, Phys.
Rev. E {\bf82}, 035205 (2010); S. Olmi, A. Politi, and A. Torcini, Europhys. Lett.{\bf 92},60007 (2010).

\bibitem{ioym2011}
I. Omelchenko, Y. Maistrenko, P. H\"{o}vel, and E. Sch\"{o}ll, Phys. Rev. Lett. \textbf{106}, 234102 (2011);
I. Omelchenko,  B. Riemenschneider, P. H\"{o}vel, Y. Maistrenko, and E. Sch\"{o}ll, Phys. Rev. E \textbf{85}, 026212 (2012);
I. Omelchenko, O. E. Omel’chenko, P. H\"{o}vel, and E. Sch\"{o}ll, Phys. Rev. Lett. \textbf{110}, 224101 (2013).

\bibitem{sheeba09}
Jane H. Sheeba, V. K. Chandrasekar and M. Lakshmanan,  Phys. Rev.  {\bf E 79}, 055203 (4 pp) (R) (2009);Phys. Rev.  {\bf E 81}, 046203 (10 pp) (2010).


\bibitem{uji2013}
S. R. Ujjwal, and R. Ramaswamy, Phys. Rev. E \textbf{88}, 032902 (2013);
D. Pazo, and E. Montbri\'o, Phys. Rev. X, \textbf{4}, 011009, (2014);
L. Schmidt, K. Sch\"{o}nleber, K. Krischer, and V. Garci'aMorales, Chaos {\bf24}, 013102 (2014).

\bibitem{chan15}
V. K. Chandrasekar, R. Suresh D.V. Senthilkumar and M. Lakshmanan, Euro.Phys Letters{\bf 111} 6008~(2015); Bidesh K. Bera, Dibakar Ghosh, M. Lakshmanan, Phys. Rev E {\bf 93}, 012205 (2015).

\bibitem{tinsley2012}
M.~R.~Tinsley, S.Nkomo and S.~Showalter, Nat. Phys. {\bf8}, 662 (2012);
S. Nkomo, M. R. Tinsley, and K. Showalter, Phys. Rev. Lett. {\bf110}, 244102 (2013).

\bibitem{Hage2012}
A. M. Hagerstrom, T. E. Murphy, R. Roy, P. H\"{o}vel, I. Omelchenko, and
E. Sch\"{o}ll, Nature Phys. {\bf8}, 658–661, (2012).

\bibitem{martens2013}
E. A. Martens, S. Thutupalli, A. Fourri\'ere and O. Hallatschek, Proc. Nat. Acad. Sci. USA, {\bf110}, 10563–10567 (2013).


\bibitem{Filatrella2008}
G.~Filatrella, A.~H.~Nielsen, and N.~F.~Pedersen, Eur. Phys. J. B {\bf 61 (4)},485 (2008).


\bibitem{rottenberg2000}
N.~C.~Rottenberg, C.~J.~Amlaner, and S.~L.~Lima, Neurosci Biobehav Rev. {\bf24},817 (2000).



\bibitem{Olb11}
 E. Olbrich, J. C. Claussen, and P. Achermann, Phil.Trans.R. Soc. A {\bf 369} ,3884 (2011)

\bibitem{Sethia:13}
G. C. Sethia, A. Sen, and G. L. Johnston, Phys. Rev. E. {\bf88}, 042917 (2013).

\bibitem{Gopal1:14}
V. K. Chandrasekar, R. Gopal,  A. Venkatesan, and M. Lakshmanan, Phys. Rev. E. \textbf{90}, 062913 (2014).



\bibitem{Sethia1:14}
G. C. Sethia and A. Sen, Phys. Rev. Lett. {\bf112}, 144101 (2014);
A. Yeldesbay, A. Pikovsky, and M. Rosenblum, Phys. Rev. Lett. \textbf{112}, 144103 (2014).


\bibitem{zak2014}
A.~Zakharova,M.~Kapeller and E.~Sch\"oll Phys. Rev. Lett {\bf112}, 154101 (2014); A. Zakharova, M. Kapeller, and E.~Sch\"oll, J. Phys. Conf.Series (2015), arXiv 1503.03371; 

\bibitem{prema:15}
K. Premalatha, V. K. Chandrasekar, M. Senthilvelan and M. Lakshmanan, Phys. Rev. E {\bf 91}, 052915 (2015);arXiv:1511.07220 (Phys. Re. E (2016) to appear). 

\bibitem{Seme2015}
N. Semenova, A. Zakharova, V. Anishchenko, and E. Sch\"oll, arXiv:1512.07036v1 (2015).




\bibitem{Prasad:05}
J. M. Cruz, J. Escalona, P. Parmananda, R. Karnatak, A. Prasad, and R. Ramaswamy, Phys. Rev. E {\bf81}, 046213 (2010); A. Prasad, J. Kurths, S. K.Dana and R. Ramaswamy, Phys. Rev. E \textbf{74}, 035204(2006);A. Sharma, M.~D.~Shrimali, A.~Prasad, R.~Ramaswamy and U.~Feudel, Phys. Rev. E {\bf84}, 016226 (2011).


\bibitem{Ammann1998} 
H. Ammann, R. Gray, I. Shvarchuck, and N. Christensen, Phys. Rev. Lett. {\bf 80}, 4111 (1998);
W. H. Zurek, S. Habib, and J. P. Paz, Phys. Rev. Lett. {\bf 70}, 1187 (1993);
S. A. Gurvitz, L. Fedichkin, D. Mozyrsky, and G. P. Berman, Phys. Rev. Lett. {\bf91}, 066801 (2003);
S. Braig and K. Flensberg, Phys. Rev. B {\bf 68}, 205324 (2003).

\bibitem{kuz2004}
A. Kuznetsov, M. Kærn, and N. Kopell, SIAM J. Appl. Math.{\bf 65}, 392 (2004);
R. Wang and L. Chen, J. Biol. Rhythms{\bf 20}, 257 (2005);
D. Gonze, S. Bernard, C. Waltermann, A. Kramer, and H. Herzel, Biophys. J.{\bf 89}, 120 (2005).

\bibitem{li2012}

B. W. Li, C. Fu, H. Zhang and X. Wang, Phys. Rev. E \textbf{86},046207 (2012). 

\bibitem{Ullner:07}
E. Ullner, A. Zaikin, E. I. Volkov, and J. Garcia-Ojalvo, Phys. Rev. Lett. {\bf 99}, 148103 (2007);
E. Ullner, A. Koseska, J. Kurths,  E. Volkov, H. Kantz, and  J. Garcia-Ojalvo, Phys. Rev. E. {\bf 78}, 031904 (2008).

\bibitem{Ambi:10}
V. Resmi, G. Ambika, and R. E. Amritkar, Phys. Rev. E \textbf{81}, 046216 (2010).

\bibitem{Amit:12}
A. Sharma, M. D. Shrimali and S. K. Dana, Chaos \textbf{22}, 023147 (2012).

\bibitem{baner:14}
T.~Banerjee and D.~Ghosh, Phys. Rev. E {\bf89}, 052912 (2014).


\bibitem{sch:15}
L. Schmidt and K. Krischer, Phys. Rev. Lett. {\textbf 114}, 034101 (2015)

\bibitem{sch1:15}
L. Schmidt and K. Krischer, Chaos {\textbf 25}, 064401 (2015).

\bibitem{bolo:16}
M. I. Bolotov, G. V. Osipov, and A. Pikovsky, Phys. Rev. E {\textbf 93}, 032202 (2016).

\bibitem{Pecora:98}
L. M. Pecora and T. L. Carroll, Phys. Rev. Lett. {\textbf 80}, 
2109 (1998).

\bibitem{asrea2015}
S. Acharyya and R. E. Amritkar, Phys. Rev. E {\textbf 92}, 052902 (2015).

\bibitem{manrubia:04}
S. C. Manrubia, A.S. Mikfailov, O. D. Zanatte {\em Emenrgence of Dynamical Orer: Synchronization Phenomena in Complex Systems} (World Scirtific, Singapore, 2004).

\bibitem{okuda:93}

K. Okuda, Physica D {\textbf 63}, 424 (1993); V.~Hakim and W.-J.~Rappel, Phys. Rev. A {\textbf 46}, R7347 (1992); N. Nakagawa and Y. Kuramoto, Prog. Theor. Phys. {\textbf 89}, 313 (1993);
Y. Kuramoto, Prog. Theor. Phys.{\textbf 94}, 321 (1995). 

\bibitem{zanette:98}
D. H. Zanette and A. S. Mikhailov, Phys. Rev. E {\textbf 57}, 276 (1998); D. H. Zanette and A. S. Mikhailov, Phys. Rev. E {\textbf 62}, R7571(R) (2000);W. Wang, I. Z. Kiss, and J. L. Hudson, Chaos {\textbf 10}, 248 (2000); M. Wickramasinghe and I. Z. Kiss, PLoS ONE {\textbf 8}, e80586 (2013).









\end{thebibliography}
\end{document}